\documentclass[3p,times,twocolumn]{elsarticle}

\usepackage{ecrc}

\volume{00}

\firstpage{1}

\journalname{Nuclear Physics B Proceedings Supplement}

\runauth{}

\jid{nuphbp}

\jnltitlelogo{Nuclear Physics B Proceedings Supplement}

\usepackage{amssymb}

\usepackage[figuresright]{rotating}

\newcommand{\gray}{$\gamma$-ray}
\newcommand{\fermilat}{\emph{Fermi}-LAT}
\newcommand{\fermi}{\emph{Fermi}}

\newcommand{\hi}{H~{\sc i}}
\newcommand{\hii}{H~{\sc ii}}
\newcommand{\ww}{0.46\textwidth}
\newcommand{\galprop}{GALPROP}

\begin{document}

\begin{frontmatter}



\dochead{}

\title{Cosmic Rays in the Milky Way and Beyond}


\author{Igor V.~Moskalenko}

\address{Hansen Experimental Physics Laboratory \& Kavli Institute for Particle Astrophysics and Cosmology, \\Stanford University, Stanford, CA 94305, U.S.A.}

\begin{abstract}
Cosmic rays (CRs) are the probes of the deep space. They allow us to study particle acceleration, 
chemical composition of the interstellar medium, and global properties of our Galaxy. 
However, until recently studies of CRs were similar to astronomical observations with blurred lenses that capture only the integral 
brightness of all stars in the field of view.
Thanks to the recent technological developments, our ``lenses" are now capable of capturing sharp images and making precise
measurements of all CR species. We have a full range of unique instrumentation for direct measurements of
CRs in space and for multi-wavelength observations of their emissions
and more missions are coming. The unveiling picture is astonishing. This paper gives a short overview of very exciting 
developments in astrophysics of CRs in the Milky Way and in other star-forming galaxies.

\end{abstract}


\end{frontmatter}


\section{Introduction} \label{intro}

After hundred years of intensive research on cosmic rays (CRs) we have learned a lot about them.
CRs present a unique resource that is not fully exploited yet. First of all, these are the only samples of 
interstellar and intergalactic matter available to us for direct probes. Information about sources of CRs
and their propagation history is encrypted in their composition. The spectra of individual CR species
can tell a story about the processes of particle acceleration and global Galactic properties. They may also bear
signatures of exotic physics we have yet to uncover. The emissions
produced by CRs in the interstellar space in the Milky Way and in other galaxies can be used to trace the density of 
CR protons and electrons across the local universe. Finally, their energy range spans 14 orders of magnitude 
from 10$^6$ to 10$^{20}$ eV providing us with an opportunity to study particle interactions at an energy
unreachable on man-made machines but only on the accelerator built by the genius of nature.

We are extremely lucky to live in this exciting time when many breakthroughs become possible thanks to 
the technological advances in measuring techniques. These new techniques have also brought new puzzles,
such as the unexpected rise in the positron fraction discovered by PAMELA \cite{2009Natur.458..607A}, 
confirmed by \emph{Fermi}-LAT \cite{2012PhRvL.108a1103A}, and measured with greater
precision by AMS-02 \cite{2013PhRvL.110n1102A}, 
the controversy of the break in proton and He spectra seemed to be well established through combined measurements by 
PAMELA \cite{2011Sci...332...69A}, ATIC \cite{2009BRASP..73..564P}, and CREAM \cite{2010ApJ...714L..89A}, 
and not confirmed by AMS-02 \cite{Ting2013}, 
a well-established difference in the spectral indices of protons, He, and heavier nuclei
(PAMELA, ATIC, CREAM, AMS-02), a flatter than expected the all-electron spectrum measured by 
\emph{Fermi}-LAT \cite{2010PhRvD..82i2004A}, PAMELA \cite{2011PhRvL.106t1101A}, and AMS-02 \cite{Ting2013},
measurements of antiprotons up to 200 GeV by PAMELA \cite{2010PhRvL.105l1101A}, and B/C ratio by PAMELA \cite{Carbone2013} and AMS-02 \cite{Ting2013}. 
In $\gamma$-rays, an all-sky survey by the \emph{Fermi}-LAT brought new advances and new puzzles in the GeV domain: 
huge ``\emph{Fermi} Bubbles" \cite{2010ApJ...724.1044S} emanating from or projected onto the Galactic center, an excess of the diffuse emission in the 
outer Galaxy \cite{2012ApJ...750....3A}, controversial claims of the detection of $\gamma$-ray lines in the 100 GeV range from several 
regions of the inner Galaxy 
\cite{2012JCAP...08..007W,2013JCAP...01..029F,2013PDU.....2...90B,2012arXiv1209.4548H,2013arXiv1303.2733B},
spatially and spectrally resolved $\gamma$-ray emission from SNR and their vicinity by \emph{Fermi}-LAT 
\cite{2013Sci...339..807A,2012ApJ...744...80A} and Cherenkov Telescopes \cite{2006ApJ...636..777A}.
Meanwhile, Voyager 1, 2 launched in 1977 that rely on the technology of the beginning of the space era are continuing their spectacular journey
to the boundary of the solar system. They appear to begin to see the true interstellar spectrum of CRs \cite{Voyager2013} which shows no gradient.  
Most of these discoveries are less that five years old and more discoveries are in the queue.

\section{CRs in the Milky Way} \label{cr}

Earlier studies of CRs were only possible through the observations of the extensive air showers in the atmosphere 
with the ground-based technique. Later on it has become possible to measure CRs directly by instruments launched into space.
These studies established that the spectrum of CRs is an almost featureless power-law with an index close to --3 extending up to
$10^{20}$ eV. The only features observed were two breaks, the steepening of the spectrum between $10^{15}$ eV and $10^{16}$ eV (the knee),
and the flattening at around $10^{18}$ eV (the ankle). The spectrum was predicted to cut off at $\sim$$10^{20}$ eV due to
the photodisintegration of nuclei and photopion production on CMB photons, a so-called Greisen-Zatsepin-Kuzmin (GZK) cutoff.

Observations at relatively low energies can be done from space and provide the most detailed information on 
CR composition including isotopic abundances and spectra of individual CR species \cite{1990A&A...233...96E,2001SSRv...99...15W,2013ApJ...770..117L}. 
This information is often used
to derive the global properties of the Galaxy, such as the diffusion coefficient and the size of the region filled with CRs (the halo),
which are then extrapolated to higher energies. The most often used are the B/C 
(secondary-to-primary) \cite{1990A&A...233...96E,2013ApJ...770..117L} and $^{10}$Be/$^9$Be 
\cite{1998ApJ...501L..59C,2002ApJ...568..210W,2006AdSpR..38.1558D} ratios as they
are best measured in CRs and their production through fragmentation of heavier nuclei is also defined better than for
many other species. However, the parameters derived from these ratios are model-dependent and can vary significantly.

The widely used in the past a so-called Leaky-Box Model \cite{1987ApJS...64..269G,1993ApJ...402..188W}, 
which treats the Galaxy as a box with uniform distributions of gas, the photon number 
density, and CR sources, is very simple where the path length distribution, an empirical functional form, is derived from
secondary-to-primary nuclei ratios. This model was appropriate for interpretation of CR data at the early stages of CR science because of the lack
of data on the distribution of CRs in the Galaxy. Meanwhile, even at that early stage the proper disk-halo diffusion model \cite{1976RvMP...48..161G} 
was put forward as a physical alternative to a non-physical uniform box. 

Observations of the diffuse emission from the Galactic plane with the first $\gamma$-ray telescopes launched into space (OSO-3, SAS-3, COS-B)
and then the first all-sky $\gamma$-ray skymap delivered by the EGRET \cite{1997ApJ...481..205H} was a clear dismissal of the uniform model. 
This was a turning point when it became clear that a further progress is impossible without a proper diffusion model,
which would model CR propagation and the diffuse emission in the whole Galaxy self-consistently.

\section{The GALPROP model for CR propagation and associated diffuse emissions} \label{galprop}

The \galprop{} project~\cite{1998ApJ...493..694M,1998ApJ...509..212S} began in late 1996 and
has now 17 years of development behind it\footnote{http://sciencewatch.com/dr/erf/2009/09octerf/09octerfStronET/}. 
The key concept underlying the \galprop{} model is that various 
kinds of data, e.g., direct CR measurements including primary and secondary nuclei, 
electrons and positrons, \gray{s}, synchrotron radiation, and so forth, are all related to the 
same astrophysical components of the Galaxy and hence have to be modeled self-consistently~\cite{1998A&A...338L..75M}.
The goal is for GALPROP-based models to be as realistic as possible and to make use of available astronomical 
information, nuclear and particle data, with a minimum of simplifying assumptions.
A complete description of the rationale and motivation is given in the review~\cite{2007ARNPS..57..285S}.

The code, originally written in fortran90, was made public in 1998. 
A C++ version was produced in 2001, and the most recent public version is v.54, which was significantly updated since its first 
release~\cite{2011CoPhC.182.1156V}. The \galprop{} code is available from a dedicated website\footnote{http://galprop.stanford.edu \label{footnote_galprop}} 
where a 500+ core facility for users to run the 
code via online forms in a web-browser is also provided \cite{2011CoPhC.182.1156V}.

\begin{figure}[tb]
\center{
\vspace{-1mm}
\includegraphics[width=\ww]{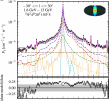}\vspace{-1.7mm}
\includegraphics[width=\ww]{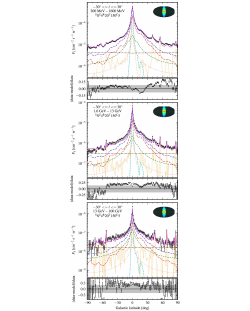}
}
\caption{\footnotesize
Latitude profile for model $^SS^Z4^R20^T150^C5$ showing the innermost $60^\circ$ about the Galactic center.
The emission components: \hi\ (red, long-dashed), H$_2$ (cyan, dash-dotted), and \hii\ (pink, long-dash-dash-dotted), and also inverse Compton (green, dashed). 
Also shown are the isotropic component (brown, long-dash-dotted), the detected sources (orange, dotted), total diffuse emission (blue, long-dash-dashed), and total model (magenta, solid). \fermilat{} data are shown as points with statistical error bars and the systematic uncertainty in the effective area is shown as a gray band.
\label{fig2}
} 
\end{figure}

\begin{figure}[tb]
\center{
\includegraphics[width=\ww]{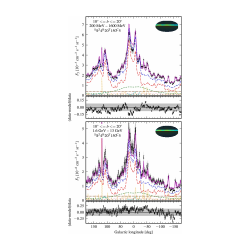}
}
\caption{\footnotesize
Longitude profile for model $^SS^Z4^R20^T150^C5$ showing north intermediate latitudes. See Figure~\ref{fig2} for legend.
\label{fig3}
} 
\end{figure}

\section{Diffuse Galactic emission} \label{diffuse}

A majority of observed \gray{s} (about 80\%) are diffuse, originating in energetic CR interactions with the interstellar gas and radiation field or 
are attributed to an ``isotropic" (presumably extragalactic) component. The large photon statistics collected by the \fermilat{} allows for a detailed study of the 
diffuse emission and the underlying spectra of CRs at distant locations, but doing it in practice is very challenging.

\begin{figure}[tb]
\center{
\includegraphics[width=\ww]{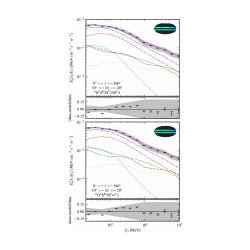}
}
\caption{\footnotesize
Spectrum of the intermediate latitude region for model $^SO^Z8^R30^T\infty^C2$ along with
isotropic background (brown, long-dash-dotted) and the detected sources (orange, dotted).
The emission components: $\pi^0$-decay (red, long-dashed), inverse Compton (green, dashed),
and bremsstrahlung (cyan, dash-dotted). Also shown is the total diffuse emission (blue, long-dash-dashed),
and the total emission including detected sources and isotropic background (magenta, solid). The \fermilat{}
data are shown as points. The gray region represents the systematic error in the \fermilat{} effective area. 
\label{fig4}
} 
\end{figure}

A recently published extensive study of the Galactic diffuse emission based on \fermilat{} data 
\cite{2012ApJ...750....3A} provides an illustrative example of the GALPROP capabilities.
A grid of 128 models covering the plausible confinement volume, source distributions, \hi\ spin temperature, and the $E(B-V)$ magnitude cuts has been explored. The resulting model skymaps were then compared with the LAT data using the maximum likelihood, the process being iterated since model parameters depend on the outcome of the fits. Models include all components of the gas (H$_2$, \hi, \hii, and the ``dark gas" through the dust reddening-corrected column density), inverse Compton from the modelled infrared and stellar photon fields (and the microwave background), bremsstrahlung, point sources, and isotropic emission. The agreement with data spanning many orders of magnitude in intensity is good. Illustrative examples are shown in Figures \ref{fig2}-\ref{fig4}; see 
\cite{2012ApJ...750....3A} for a definition of the model coding. 

\begin{figure}[tb]
\center{
\includegraphics[width=\ww]{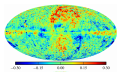}
}
\caption{\footnotesize
A fractional residual map, (model -- data)/data, in the energy range 200 MeV -- 100 GeV. Shown are residuals for model $^SS^Z4^R10^T150^C5$.
The map has been smoothed with a $0^\circ\!\!\!.5$ hard-edge kernel.  
\label{fig5}
} 
\end{figure}

The agreement with all available data (direct measurements of CRs, diffuse $\gamma$-ray and synchrotron emission) is impressive and is a proof that
the basic features of CR propagation are reproduced correctly.   
However, the fits are not perfect. 
Discrepancies between the physical model and high-resolution data (Figure \ref{fig5}), the residuals, are a potential gold mine of new phenomena. 
Every extended source and/or process that is not included in the model pops up and exposes itself as a residual. One example is the \fermi{} Bubbles above and below the plane that were associated with activity at the Galactic center \cite{2010ApJ...724.1044S}. The other residuals, including the negative ones, are equally interesting.

\section{The positron fraction} \label{positrons}

The origin of the unexpected rise in the positron fraction \cite{2009Natur.458..607A,2012PhRvL.108a1103A,2013PhRvL.110n1102A} is not understood yet.
The conventional propagation models \cite{1998ApJ...493..694M} predict the secondary positron fraction 
(positrons produced by CRs interacting with interstellar gas) 
to fall with energy while the data clearly shows that it rises and the discrepancy is very significant. 
The two dominant interpretations, in terms of the dark matter annihilation or a pulsar contribution, each has its own
difficulties. The dark matter hypothesis has to explain the amount of the positron excess, the lack of a similar excess
in antiprotons, and the lack of a clear signal in $\gamma$-rays; see \cite{2009PhRvD..79a5014A} for a viable explanation,
although it requires a WIMP in the TeV mass range. 
The pulsar hypothesis (e.g., \cite{2009JCAP...01..025H}) is based on a single assumption 
that pulsars are producing (primary) electrons and positrons in equal numbers, but lacks a convincing 
calculation of the number of ejected particles and their spectrum. 

There are also several papers claiming that the rising positron fraction is consistent with secondary origin.
None of them is convincing so far, here is why.

Papers \cite{2010MNRAS.405.1458K,2011JCAP...11..037B,2013arXiv1305.1324B} derive an upper limit 
for the positron fraction assuming no inverse Compton energy losses. The difference between the predicted and
observed fraction is interpreted as due to the energy-dependent residence $t_{esc}$ time of positrons in the Galaxy. 
It should be shorter than the cooling time $t_{cool}$ at high energies and longer than $t_{cool}$ at low energies:
\begin{eqnarray}
t_{esc}(E/Z=10~{\rm GeV}) &\ge& t_{cool} (E=10~{\rm GeV}) \nonumber\\
&\sim& 30~{\rm Myr}~\left(\bar{U}_T\over1~{\rm eV~cm}^{-3}\right)^{-1}, \label{a}\\
t_{esc}(E/Z=200~{\rm GeV}) &\le& t_{cool} (E=200~{\rm GeV}) \nonumber\\
&\sim& 1.7~{\rm Myr}~\left(\bar{U}_T\over1~{\rm eV~cm}^{-3}\right)^{-1}.\label{b}
\end{eqnarray}
Here $\bar{U}_T$ is the time averaged total electromagnetic energy density in the propagated region. 
A consistency with the effective grammage $X_{esc}=8.7 \left([E/Z]/10~{\rm GeV}\right)^{-0.5}$ g cm$^{-2}$, derived from the B/C ratio, requires
that the mean number density of the gas traversed by CRs to be energy dependent. 

This interpretation has a problem with the primary nuclei. Eqs.~(\ref{a}), (\ref{b}) allow one to derive the average
index of the energy dependence for $t_{esc}$:
\begin{equation}
\delta\le{{\log\left(30~{\rm Myr}/ 1.7~{\rm Myr}\right)}\over{\log\left(10~{\rm GeV}/200~{\rm GeV}\right)}}\approx-0.96. 
\end{equation}
Such a fast escape would require the injection spectrum to have an index $\beta=\alpha-\delta\ge-1.7$, where $\alpha\approx-2.7$ is the measured 
spectral index of CRs, while the observations of $\gamma$-ray
emission from SNRs \cite{2013Sci...339..807A,2012ApJ...744...80A,2006ApJ...636..777A} require it to be rather steep $\beta<-2$.
Besides, if such a fast escape is extrapolated to higher energies, it would result in a large CR anisotropy, inconsistent with observations
(e.g., \cite{2007ARNPS..57..285S}).

A paper \cite{2009PhRvL.103k1302S} looks into the effect of inhomogeneity of CR sources, in particular their concentration
in spiral arms. This may help to reproduce the observed spectrum of CR electrons because it is dominated by the primary electrons,
which are affected by significant inverse Compton losses at high energies and could come only from nearby sources.
Regarding secondary positrons, the authors write: ``Protons are not affected by cooling and are therefore distributed rather smoothly in the 
galaxy even if their sources are inhomogeneous. The secondary positrons (that are produced by smoothly distributed protons)
are only weakly affected by the inhomogeneity of the sources..." Obviously, it is the theoretically predicted concave shape of the 
electron spectrum that makes the positron fraction to rise. If we disregard the theoretical prediction of the electron spectrum and
stick with the \fermilat{} \cite{2010PhRvD..82i2004A} and PAMELA \cite{2011PhRvL.106t1101A} data, 
the best we can get is an almost flat positron fraction inconsistent with AMS-02 data \cite{Ting2013}. 

Papers \cite{2010PhRvD..82b3009C,2013arXiv1305.1242C} propose that the grammage traversed by CRs in the Galaxy
is fairly small $\approx$1.7 g cm$^{-2}$ and energy independent. In this Nested Leaky-Box model,
the energy dependent (and dominating) part of the grammage, 
necessary to reproduce the observed shape of the B/C ratio, is traversed near the CR sources. 
Such interpretation predicts a constant or very flat B/C ratio above $\sim$100 GeV/nucleon while
PAMELA \cite{Carbone2013} and especially AMS-02 \cite{Ting2013} show that the ratio continues to fall up to $\sim$400 GeV/nucleon,
but the error bars are still large. 
Besides, if most of the total grammage below $\approx$30 GeV/nucleon is traversed near the sources, 
they will be seen as very bright GeV $\gamma$-ray sources with soft spectrum while the diffuse emission will be significantly dimmer 
than observed (see Section \ref{diffuse}). Given such a fast escape of particles from the Galaxy, it would require the SNe rate that is several times
larger than the current estimate, 3 SNe per century, to sustain the observed CR flux below $\approx$30 GeV/nucleon.

The only hypothesis \cite{2010ApJ...710..236S} that does not contradict to other CR measurements proposes that positrons are 
produced through annihilation of diffuse TeV photons on starlight, but, in turn, 
this would require unrealistically high density of starlight photons as the authors 
themselves remark in the paper.

\begin{figure}[tb!]
\centering
\includegraphics[width=\linewidth]{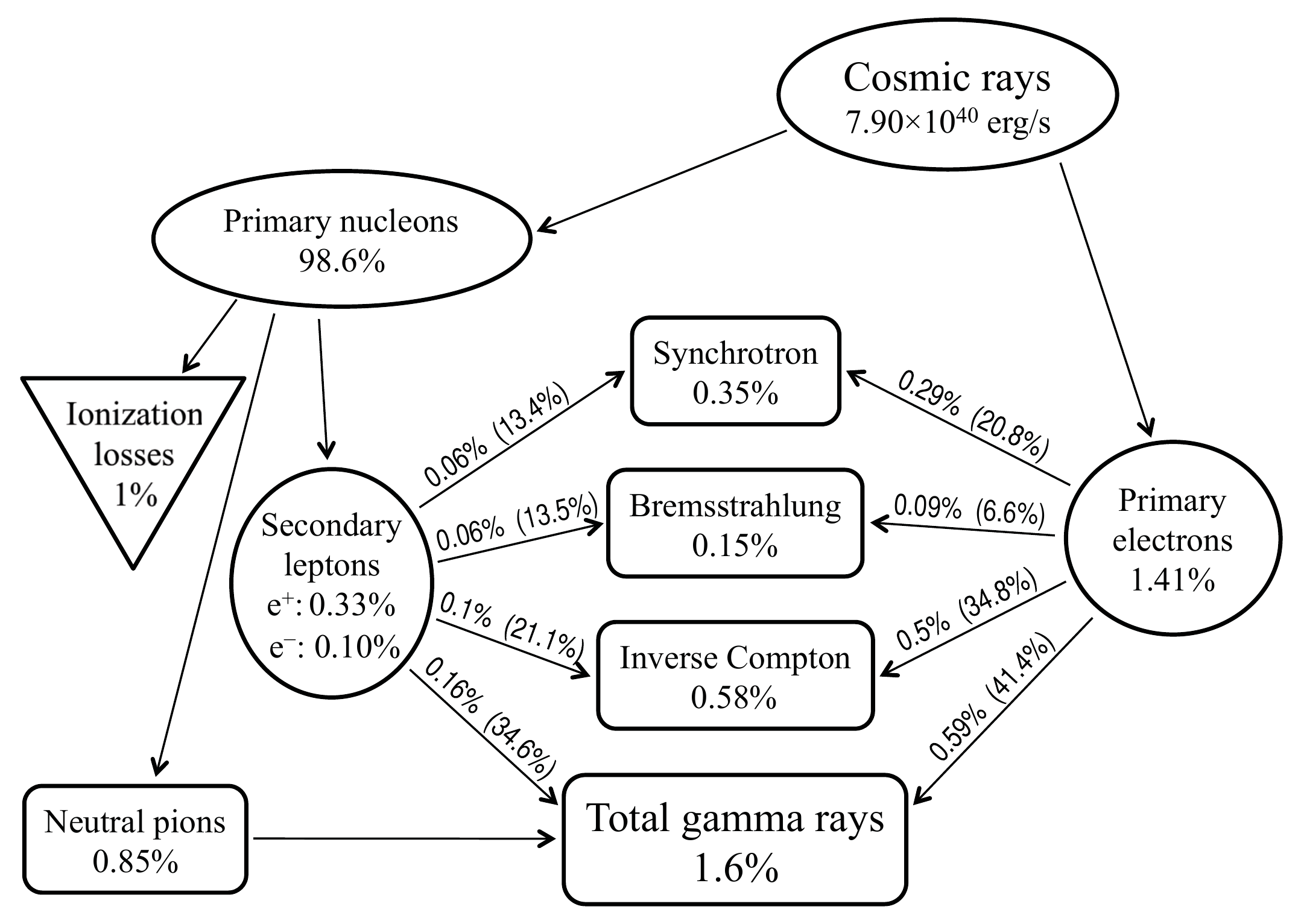}
\caption{The luminosity budget of the Milky Way galaxy 
calculated for a model with 4 kpc halo~\cite{2010ApJ...722L..58S}. The percentage figures
are shown with respect to the total injected 
luminosity in CRs. 
The percentages in brackets show the values relative to the luminosity of 
their respective lepton populations 
(primary electrons, secondary electrons/positrons).}
\label{MW}
\end{figure}

\begin{figure}[tb!]
\centering
\includegraphics[width=\linewidth]{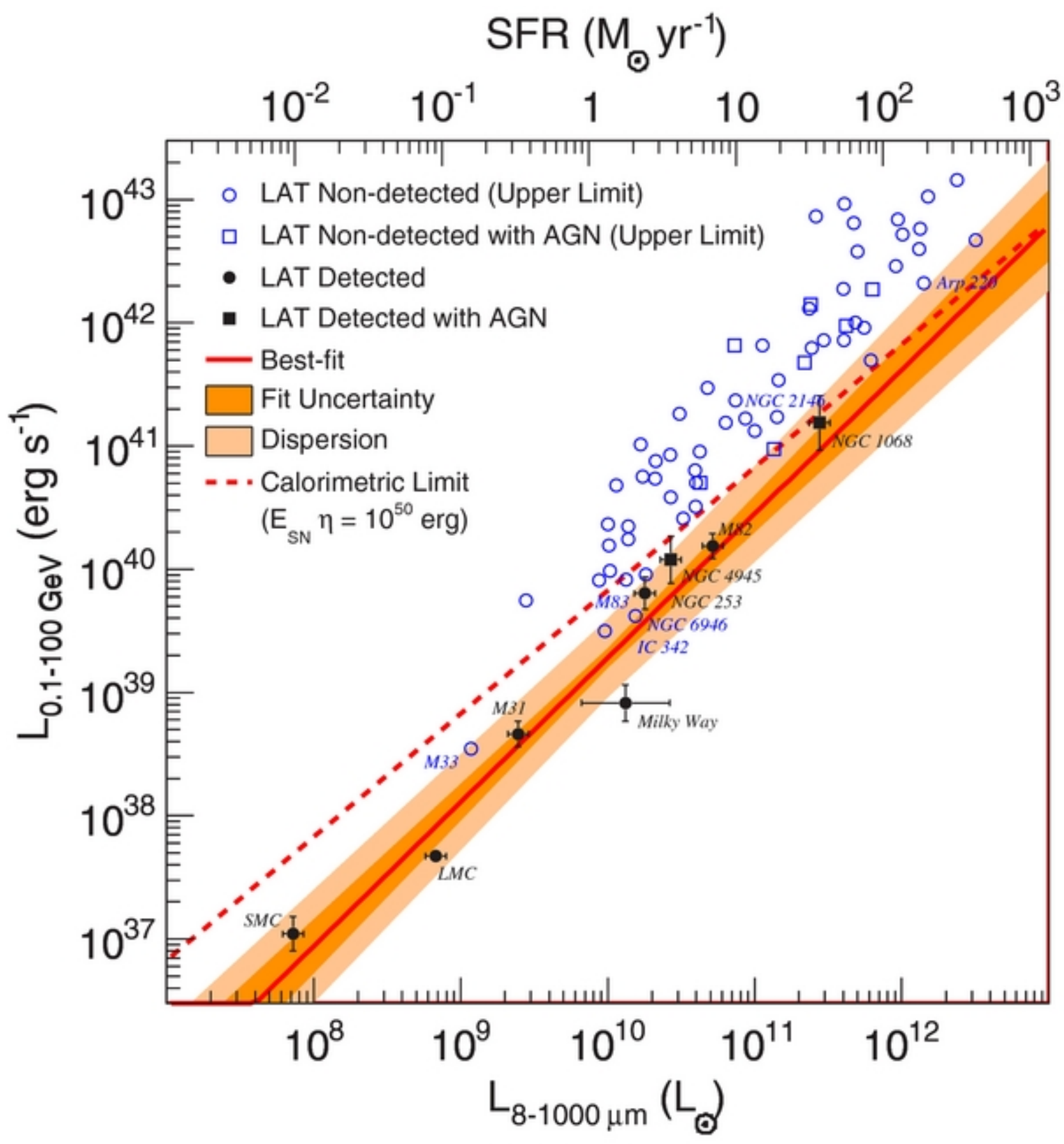}
\caption{$\gamma$-ray luminosity (0.1--100 GeV) vs.\ total IR luminosity (8-1000 $\mu$m) for star-forming galaxies.
Galaxies significantly detected by \fermilat{} are indicated with filled symbols whereas galaxies with gamma-ray flux upper limits (95\% confidence level) are marked with open symbols. The upper abscissa indicates the star-formation rate (SFR) estimated from the radio continuum luminosity. The best-fit power-law relation is shown by the red solid line along with the fit uncertainty (darker shaded region), and intrinsic dispersion around the fitted relation (lighter shaded region). The dashed red line represents the expected $\gamma$-ray luminosity in the calorimetric limit assuming an average CR luminosity per supernova of $10^{50}$ erg.}
\label{starforming}
\end{figure}

\section{Other galaxies} \label{galaxies}
Observations of the diffuse \gray{} emission from normal galaxies (LMC, SMC, M~31) and the starburst galaxies (M~82, NGC~253)
by the \fermilat~\cite{2010ApJ...709L.152A,2010A&A...512A...7A,2010A&A...523L...2A,2010A&A...523A..46A} and by the atmospheric Cherenkov telescopes~\cite{2009Sci...326.1080A,2009Natur.462..770V} show that
CRs is a widespread phenomenon associated with the process of star formation. 
The Milky Way is the best-studied non-AGN dominated star-forming 
galaxy, and the only galaxy that direct measurements of CR intensities and 
spectra are available.
However, because of our position inside, the derivation of 
global properties is not straightforward and requires detailed 
models of the spatial distribution of the emission. 
Meanwhile, understanding the global energy budget of processes related to the injection
and propagation of CRs, and how the energy is distributed across the 
electromagnetic spectrum, is essential to interpret the 
radio/far-infrared relation~\cite{Helou1985,Murphy2006}, galactic 
calorimetry~\cite{Volk1989}, and predictions of 
extragalactic backgrounds~\cite{Thompson2007,Murphy2008}, and for
many other studies.

Such calculations were carried out in~\cite{2010ApJ...722L..58S}. The luminosity spectra were
calculated for representative Galactic propagation models that are 
consistent with
CR, radio, and \gray{} data.
Figure~\ref{MW}
shows the detailed energy budget for a model corresponding to the middle
range of the plausible models.
About 1.8\% of the total CR luminosity goes into the primary and secondary
electrons and positrons, however, the IC scattering contributes half of the total \gray{} luminosity
with the $\pi^0$-decay contributing another half.
The relationship between far-infrared and radio luminosity appears to be consistent with that found for galaxies in general.
The Galaxy is found to be a good CR electron calorimeter, 
but {\it only} if \gray{} emitting processes are taken into account. 
The synchrotron emission alone accounts for only one third of the total 
electron energy losses with $\sim$10-20\% of the total synchrotron 
emission from secondary CR electrons and positrons.

Well-resolved in $\gamma$-rays the Milky Way galaxy provides an important calibration point for other star-forming 
galaxies. Figure~\ref{starforming} shows the $\gamma$-ray luminosity (0.1--100 GeV) vs.\ total IR luminosity (8--1000 $\mu$m) for 
star-forming galaxies detected by \fermilat{} \cite{2012ApJ...755..164A}. All detected galaxies are below the ``calorimetric limit," which means that their total
$\gamma$-ray luminosity is consistent with the star-formation rate (SFR) and the CR output into the interstellar medium, i.e., this emission
is ``diffuse." Going from lower left to the upper right, we transition from softer-spectra galaxies in which energy-dependent diffusive losses
are important to the harder-spectra starburst galaxies in which the energy
loss rate due to proton-proton interactions is considerably faster than the diffusion timescale.

\medskip
I am grateful to Marty Israel for many stimulating discussions on the origin of the rise in the positron fraction.
This work was supported by NASA grants NNX09AC15G and NNX13AC47G.




\nocite{*}
\bibliographystyle{elsarticle-num}
\bibliography{moskalenko}







\end{document}